\documentclass{article}
\newcommand{\Rbig}{\mbox{I\hspace{-.19em}R}}

\newcommand{\ov}{\overline}
\begin{document}
\title{Characterization of SU(1,1) coherent states in terms of affine group
 wavelets \footnote{to appear in {\it J.Phys.A}}}
\author{Jacqueline BERTRAND and Mich\`{e}le IRAC-ASTAUD \\
Laboratoire de Physique Th\'{e}orique de la mati\`{e}re
condens\'ee\\ Universit\'{e} Paris VII\\2 place Jussieu, F-75251
Paris Cedex 05, FRANCE\\ e-mail : bertrand@ccr.jussieu.fr,
mici@ccr.jussieu.fr}
\date{}
\maketitle

\begin{abstract}
The Perelomov coherent states of $SU(1,1)$ are labeled by elements
of the quotient of $SU(1,1)$ by the compact subgroup. Taking
advantage of the fact that this quotient is isomorphic to the
affine group of the real line, we are able to parameterize the
coherent states by elements of that group or equivalently by
points in the half-plane. Such a formulation permits to find new
properties of the $SU(1,1)$ coherent states and to relate them to
affine wavelets.
\end{abstract}

%%% ----------------------------------------------------------------------

%%% ----------------------------------------------------------------------
%%% ----------------------------------------------------------------------

\section{Introduction}

%%%%%%%
%%%%%%%
Coherent states associated with the affine and $SU(1,1)$ groups
have been introduced  in   different situations and have led to
applications in   fields of physics that are not directly
connected. It is the purpose of the present work to exhibit the
relations existing between some of   those states.
\par

The group of affine transformations of the real line
%arises in
%problems having an invariance under changes of units. As such, it
plays an essential role in
 the analysis of acoustic and electromagnetic signals depending on
one variable (e.g. the time). Systems of coherent states
associated with that group, more
 recently known as wavelets, have been  introduced as overcomplete bases of a
  Hilbert space
  in which a unitary irreducible representation  acts
  \cite{klauder} \cite{kaiser}. Their construction  requires
  the choice of an admissible fiducial state (or mother wavelet) which is subsequently
  displaced by the operators of the representation under consideration. For
  a special  choice of this basic state, the corresponding system
  of coherent states has  minimal properties that
   have proved useful in applications \cite{bertrand}. \par

The role of $SU(1,1)$ in physics, especially in quantum physics,
has been recognized for a long time and its coherent states  have
been extensively studied.
  Due to the more complex structure of the group,    several definitions are
   available even with the  sole  requirement of obtaining overcomplete bases  \cite{perelomov}
   \cite{barut} \cite{trifonov}.
  Restricting to   systems of coherent states generated by
   displacement of a fundamental
    state, one still obtains different solutions, depending on the group
     representation  and the initial  state.
  To be able to make a connection with the affine group coherent states,
   we will   consider only the discrete series representations acting on a
 rotation invariant basic state. As shown in \cite{perelomov}, this choice  leads
  to a system of coherent states  labeled by the elements   of the
quotient of $SU(1,1)$ by the rotation group. The study could be
adapted to the fundamental series of representations but the
fiducial state must always have a rotation invariance.
\par

  The question  of the comparison between
  %a possible relation between
  the   coherent states
  corresponding to the affine and $SU(1,1)$ groups arises
   because both  appear in the problem of
the Morse potential \cite{gerry} and, more fundamentally,  because
the affine group is isomorphic to a subgroup of $SU(1,1)$. Some
preliminary results have been obtained in \cite{cz}. In the
following, we will establish the precise relation existing between
the two sets of states and discuss the applications.
\par
  The study is most easily performed by realizing the discrete series
  representations of $SU(1,1)$ in  spaces $L^2_k(\Rbig^+)$ of functions on the
  half-line in which irreducible representations of the affine group are
  naturally realized, as is recalled in section 2. In sections 3 and 4, we
  give explicit expressions for the canonical bases   and   the Perelomov coherent
  states in these spaces. In section 5, the latter states are parameterized in
  terms of the affine group. The identification to specific affine wavelets and
  the comparison with Morse states follow. New properties of the $SU(1,1)$
  coherent states are obtained in section 6.

\section{Unitary representations of the affine and  $ SU(1,1)$ groups in spaces
$L^2_k(\Rbig^+)$}\label{group}

In this section, we recall useful formulas concerning the
isomorphic groups $ SU(1,1)$ and $SL(2,\Rbig)$ and their affine
subgroups. The group $SU(1,1)$ consists of matrices of the form:
\begin{equation}\label{matricessu11}
\Gamma = \left(
  \begin{array}{c}
    \gamma_1 \quad \gamma_2 \\
    \overline{\gamma}_2\quad \overline{\gamma}_1\\
  \end{array} \right)
\end{equation}
where $\gamma_1$, $\gamma_2$ are complex numbers such that:
\begin{equation}
   \mid \gamma_1 \mid^2 - \mid \gamma_2 \mid^2
  =1.
\end{equation}
 The generators of its Lie
algebra are:
\begin{equation}\label{gen}
  J_0 = \frac{1}{2}\left(
  \begin{array}{c}
  1\quad 0\\
  0\quad -1\\
  \end{array}
  \right), \quad
  J_1 = \frac{-i}{2}\left(
  \begin{array}{c}
  0 \quad 1\\
  1\quad 0\\
  \end{array}
  \right),\quad
   J_2 = \frac{1}{2}\left(
  \begin{array}{c}
  0 \quad -1\\
  1 \quad 0\\
  \end{array}
  \right),
\end{equation}
  and satisfy the commutation relations:
\begin{equation}\label{algebresu11}
  [J_1,J_2]=-i J_0, \quad
   [J_2,J_0]=i J_1, \quad
    [J_0,J_1]=i J_2\\
\end{equation}
These commutation relations define the abstract algebra $su(1,1)$.
In the several different realizations of this algebra considered
below, we will always denote the generators by $J_0, J_1, J_2$.
The Casimir operator, defined as $C= J_1^2+J_2^2-J_0^2$, commutes
with the three generators.
 We introduce:
\begin{equation}\label{J+-}
J_\pm = J_1 \pm i J_2
\end{equation}
In some instances, it will be more convenient to consider the
   group $SL(2,\Rbig)$ consisting of matrices:
\begin{equation}\label{sl2r}
  g=\left( \begin{array}{cc}
    g_{11} & g_{12} \\
   g_{21} & g_{22} \
  \end{array}\right)
\end{equation}
where the real numbers $ g_{11},g_{12},g_{21},g_{22}$ verify the
condition:
\begin{equation}\label{sl2r2}
   g_{11} g_{22}-g_{12}g_{21}=1
\end{equation}
The explicit form of the isomorphism between $SU(1,1)$ and
$SL(2,\Rbig)$ is given by:
\begin{equation}\label{iso}
\begin{array}{c}
   \gamma_1 = \frac{1}{2} \left[  g_{11}+ g_{22}+i(g_{12}-g_{21}) \right]
   \\[2mm]
 \gamma_2 = \frac{1}{2} \left[g_{12}+g_{21}-i( g_{22}- g_{11}) \right]
\end{array}
\end{equation}
The affine group $\mathcal{A}$ consists of elements $(a,b)$, $a>0$
and $b$ real, acting on an element $x$ of the real line according
to: $x \rightarrow ax+b$. It is isomorphic to the subgroup of
elements of $SL(2,\Rbig)$ given by:
\begin{equation}\label{mab0}
\left(
  \begin{array}{cc}
   \sqrt{a} & {\displaystyle \frac{b}{\sqrt{a}} }\\ [4mm]
    0 & {\displaystyle \frac{1}{\sqrt{a}}} \
  \end{array}
  \right)
  \quad a>0, \quad b \in \Rbig
\end{equation}
  and to the subgroup of
$SU(1,1)$ consisting of matrices $M(a,b)$ defined by:
\begin{equation}\label{mab}
  M(a , b) =\frac{1}{2\sqrt{a}}
  \left(
  \begin{array}{cc}
    a+1+ib & i(a-1-ib)\\[2mm]
    -i(a-1+ib) & a+1-ib \
  \end{array}
  \right)
%  \quad a>0, \quad b \in \Rbig
\end{equation}
The discrete series representations of $SU(1,1)$ (and
$SL(2,\Rbig)$),
  labeled by a number $k \geq 1$ such that $2k$ is an integer, are unitary and inequivalent.
They will be described in three equivalent realizations.\par

\noindent{\bf Representation ${\cal{T}}^k(\Gamma)$ in space
${\cal{H}}_z$ : }
 In the space ${\cal{H}}_z$ of functions $f(z)$ that are analytic
inside the unit circle, the operators ${\cal{T}}^k(\Gamma)$
representing the elements of the group $SU(1,1)$ are defined by:
\begin{equation}\label{groupe-z}
{\cal{T}}^k(\Gamma) f(z) = (\gamma_2 z +\overline{\gamma_1})^{-2k}
f\left(\frac{\gamma_1 z +\overline{\gamma_2}}{\gamma_2 z
+\overline{\gamma_1}}\right)
\end{equation}
This representation is unitary for the scalar product:
\begin{equation}\label{scaz}
(f,f')= \int_D \overline{f(z)} f'(z) (1-\mid z\mid^2)^{2k-2}
d\overline{z} dz ,\quad    D=\{z,\mid z\mid <1\}
\end{equation}
   \par

\noindent The generators of the algebra $su(1,1)$ are represented
by the differential operators
\begin{eqnarray}
J_0&=& z\partial_z +k \label{1jfz} \\
 J_+&=& i( z^2\partial_z +2kz) \label{2jfz}\\
 J_-&=&
-i \partial_z \label{3jfz}
\end{eqnarray}

\noindent{\bf Representation $T^k$ in space $H_w$ :}
 Another realization is more adapted to the group $SL(2,\Rbig)$. It
is realized in
 the space $H_w$ of holomorphic functions $h(w)$ on the half plane $Re(w)>0$,
  equipped with the scalar
product:
\begin{equation}\label{scaw}
(h,h')\equiv \int_{Re(w)>0} \overline{h(w)}h'(w) (Re(w))^{2(k-1)}
dwd\bar{w}
\end{equation}
The operator $T^k$ representing and element $g \in SL(2, \Rbig )$
is defined by:
\begin{equation}\label{tk}
  T^k(g)h(w) =     (ig_{21} {w} + g_{11})^{-2k} h \left(
  \frac{ g_{22}w-ig_{12}}{ig_{21}w+ g_{11}}
  \right)
\end{equation}
The space $H_w$ of functions $h(w)$ is isomorphic to the space
$\mathcal{H}_z$ of functions $f(z)$ under the following
transformation:
\begin{equation}\label{isohh}
  %f(z) = \lambda \, 2^k (-i)^{-2k} (-z+i)^{-2k} h\left(
  %\frac{{z}+i}{-{z}+i} \right)
  f(z) =  2  (iz+1)^{-2k} h\left(
  \frac{1-iz}{1+iz} \right)
\end{equation}

 \par

\noindent The generators of $su(1,1)$ in representation $T^k$ are
found to be:
\begin{equation}\label{genw}
\begin{array}{l}
  J_0 = \frac{1}{2} \left[(w^2-1)\partial_w +2kw \right]\\[2mm]
   J_1 = -\frac{1}{2} \left[(w^2+1)\partial_w +2kw \right] \\[2mm]
   J_2 =  -i(w \partial_w +k)
\end{array}
\end{equation}

\noindent{\bf Representation $U^k$ in $L^2_k(\Rbig^+)$ :} There is
another form of the representations of $SL(2,\Rbig )$
 and hence of $SU(1,1)$, that has been studied in detail in
 \cite{unterberger} and that will be essential  here. It acts in
   the Hilbert space  $L^2_k(\Rbig^+)$ of functions $\psi(y)$, $y>0$, on
 the half-line with the scalar product:
\begin{equation}\label{scay}
(\psi,\psi') = \int_0^\infty \overline{\psi(y)}\psi'(y) y^{1-2k}
dy
\end{equation}
This space is applied isomorphically into $H_w$ by a Laplace
transformation written explicitly as:
\begin{equation}\label{laplace}
 h(w)=  \sqrt{\frac{(4\pi)^{2k-1}}  {(2k-2)! } }\int_0^{\infty} \psi(y) e^{-2\pi wy} \; dy
\end{equation}
The representation $U^k(g)$ of $SL(2,\Rbig)$ in $L^2_k(\Rbig^+ )$
that is equivalent to $T^k$ can be written from there. In the
following, we will only need the explicit form of the restriction
of $U^k$ to the affine subgroup of $SU(1,1)$, which consists of
elements   $M(a,b)$ defined in (\ref{mab}). It is equal to:
\begin{equation}\label{repu}
  U^k(M(a,b))\psi(y) =  a^{(1-k)} \, e^{2i\pi by} \, \psi( ay),
\end{equation}
\noindent   This restriction is an irreducible representation of
the affine group  ${\cal{A}}$. Notice that the representations of
${\cal{A}}$ corresponding to different values of $k$ are
equivalent.\par

 \noindent The generators of the representation
$U^k$ are obtained from
 (\ref{genw}) and (\ref{laplace}):
\begin{eqnarray}
J_0&=& \frac{1}{4\pi}\left( -y\partial_y^2 + 2 (k-1)\partial_y
+4\pi ^2 y\right) \label{J-sur-f(y)} \\[2mm]
 J_1 & = -& \frac{1}{4\pi}\left( -y\partial_y^2 + 2 (k-1)\partial_y -4\pi ^2
y\right) \label{2J-sur-f(y)} \\[2mm]
 J_2  & = & i\left( y\partial_y+1-k\right) \label{3J-sur-f(y)} \
\end{eqnarray}

%%%%%%%%%%%%%%%%%%
%%%%%%%%%%%%%%%%%%

\section{Construction of the
canonical basis for the algebra $su(1,1)$} \label{repres}
  We now recall the construction of the
canonical basis for the discrete series representation of the
algebra $su(1,1)$ and give its explicit form in the spaces
${\mathcal{H}}_z$, $H_w$ and $L^2_k(\Rbig^+)$. In those
representations, the value of the Casimir operator is
 $C= k(1-k)\hat{I}$ and  the
set of normalized vectors $\mid k, m>$ is defined by:
\begin{equation}\label{represkm}
\begin{array}{lll}
J_0 \mid k, m> &= (k+m) \mid k, m>&\\[2mm] J_- \mid k, m> &=
\sqrt{[m]_k} \mid k, m-1>&\quad [m]_k \equiv m(2k+m-1) \\[2mm]
 J_+ \mid k, m> &=
\sqrt{[m+1]_k} \mid k, m+1>&\\
\end{array}
\end{equation}
where $m$ is a positive integer.
% The vector of the canonical basis are obtained by action of the
%creation operator $J_+$ on
\noindent The fundamental vector $\mid k 0 >$ is defined by :
\begin{equation}\label{1k0}
  J_0 \mid k 0> =  k \mid k 0>,   \quad
   J_- \mid k 0>  =  0
\end{equation}
The two equations are necessary so long as the representation
space is not specified. The vectors $|km>$ are constructed in
terms of $\mid k 0 >$ as:
\begin{equation}\label{km}
\mid k m> = \frac{1}{\sqrt{[m]_k!}} (J_+)^m\mid k 0>, \quad [m]_k!
\equiv \prod_{i=1}^m [i]_k = \frac{m!(2k+m-1)!}{(2k-1)!}
\end{equation}
{\bf{Canonical basis in ${\cal{H}}_z$} :}
%\subsection{Construction of the canonical basis in ${\cal{H}}_z$}
%
Using the construction previously described, we obtain :
\begin{equation}\label{km-z}
<\overline{z}\mid k m> =\sqrt{ \frac{(2k-1)
}{\pi}}\frac{\sqrt{[m]_k!}}{m!}(iz)^m
\end{equation}
{\bf{Canonical basis in $H_w$ :}}
The normalized
   states of the canonical basis in $H_w$ are obtained from the
   inverse of
transformation (\ref{isohh}). They are equal to
\begin{equation}\label{km-w}
<\overline{w}\mid km> =  \sqrt{ \frac{(2k-1) }{\pi}}
\frac{\sqrt{[m]_k!}}{m!}\;2^{2k-1}\frac{ (1-w)^n}{(w+1)^{2k+n}}
\end{equation}
{\bf{Canonical basis in $L^2_k(\Rbig^+)$}:}
In the space $L^2_k(\Rbig^+)$, the vectors of the canonical basis
are the Laplace transforms of the previous ones. But the easiest
way to obtain them is by a direct construction using the
  explicit expressions of the generators
(\ref {J-sur-f(y)})-(\ref{3J-sur-f(y)}). The fundamental vector
$<y\mid k 0
>$ is defined again by conditions (\ref{1k0})
which are written in space $L^2_k(\Rbig^+)$ as two compatible
differential equations that
 reduce to
\begin{equation}\label{k0-y-2}
\left(y\partial_y +2\pi y -2k+1\right)<y\mid k 0 > =0
\end{equation}
%
% We easily obtain
 The solution of (\ref{k0-y-2}) normalized for  the scalar
product (\ref{scay}), is:
\begin{equation}\label{k0-y}
<y\mid k 0 > = \frac{(4\pi)^k  }{\sqrt{(2k-1)!}}\;
y^{2k-1}\exp(-2\pi y)
\end{equation}
and verifies both equations (\ref{1k0}).
 Substituted in(\ref{km}), this expression leads to:
 \begin{equation} \label{2kmy}
 <y\mid k m >=\frac{(4\pi)^k  }{\sqrt{(2k-1)![m]_k!}}y^{2k-1} \exp(-2\pi y)
\times P_m(y)
\end{equation}
\noindent where $P_m(y)$ are polynomials of degree $m$ in $y$.
 These polynomials
 %$P_m(y)$
    satisfy the following two  equations that
result from the action of the $su(1,1)$ generators, expressed in
(\ref{J-sur-f(y)})-(\ref{3J-sur-f(y)}), on $<y|km>$~:
\begin{equation}\label{rec4-Pm}
\left(y\partial_y + 2k+m-4\pi y
 \right) P_m (y)=- P_{m+1}(y)
\end{equation}
%
%and
%
\begin{equation}\label{rec5-Pm}
\left(-y\partial_y + m
 \right) P_m (y)=- [m]_k P_{m-1}(y)
\end{equation}
These relations with the initial condition
\begin{equation}\label{pzero}
P_0(y)=1
\end{equation}
%
%define the polynomials $P_m(y)$
lead to the expression of the polynomials $P_m(y)$:
\begin{equation}\label{laguerre}
P_m(y) = (-1)^m m! \, L_m^{2k-1}(4\pi y)
\end{equation}
where $L_m^{2k-1}$ are the Laguerre polynomials \cite{gradshteyn}.

%%%%%%%
\section{$SU(1,1)$ coherent states in Perelomov's parameterization} \label{pere}
These coherent states are generated by action of the following
elements of the group $SU(1,1)$:
\begin{equation}\label{exiji}
 %\mid \zeta> \equiv
 e^{\xi J_+ - \bar{\xi} J_-}  \quad \left(
   \xi \equiv  \frac{\tau}{2}e^{-i\varphi}  \quad \tau \in
  \Rbig  \quad 0 \leq \varphi < 2\pi \right)
 \end{equation}
on the fundamental state. The result is:
\begin{equation}
 \mid \zeta >  \equiv e^{\xi J_+ - \bar{\xi} J_-}\mid k 0>= (1-\mid
\zeta\mid^2)^k \sum_{m\geq 0}\sqrt{[m]_k!} _,
\frac{\zeta^m}{m!}\mid k m> \label{2coh-per1}
\end{equation}
\noindent where
\begin{equation}\label{zeta}
  \zeta=\tanh{\frac{\tau}{2}}\exp(-i\varphi )
\end{equation}

 Since $\mid k 0>$ is an eigenstate
of $J_0$,
%Because of condition (\ref{1k0}),
  the set of coherent states will depend only on the quotient
of $SU(1,1)$ by the rotation group ${\mathcal{R}}$. Such a
quotient is isomorphic to the upper sheet of the hyperboloid
$n_0^2-n_1^2 -n_2^2=1$   parameterized by $(\tau, \varphi)$  in
the following way:
\begin{equation}\label{hyp}
  \vec{n} = (\cosh \tau,\sinh\tau\cos\varphi,
 \sinh\tau\sin\varphi)
\end{equation}
and to its stereographic projection    onto the inside of the unit
disk parameterized by $\zeta$ given in (\ref{zeta}).
\par

The coherent states thus obtained verify the completeness
relation:
\begin{equation}\label{rel-fer-zeta}
\frac{2k-1}{\pi}\int_D \frac{d^2\zeta}{(1-\mid \zeta
\mid^2)^2}\mid \zeta><\zeta\mid =1 , \quad D=\{\zeta,\mid \zeta
\mid < 1\}
\end{equation}
and form an overcomplete set. The whole set of rays defined by the
coherent states $|\zeta >$ is stable under action of $SU(1,1)$. In
particular,  the rotation subgroup acts on such a   state  through
the operator $  \exp(-i \theta J_0) $ as:
% The
%property of $\mid k m>$ to be an eigenvector of $J_0$ leads to the
%expression:
 % a coherent state
 %multiplied by a phase factor namely :
\begin{equation}\label{groupe-CS}
\exp(-i \theta J_0)  \mid \zeta> =  e^{-ik\theta  }\, \mid
 \zeta e^{-i\theta}>.
\end{equation}
\noindent The explicit form of the Perelomov coherent states in
the different spaces considered above results from the expressions
(\ref{km-z}), (\ref{km-w}) and (\ref{2kmy}) of the canonical
basis.
\par
\noindent{\bf Coherent states in ${\cal{H}}_z$}:
\begin{equation}\label{per-CS-z}
<\overline{z} \mid \zeta>= \sqrt{\frac{2k-1}{\pi}}\frac{(1-\mid
\zeta\mid^2)^k}{(1-i\zeta z)^{2k}}
\end{equation}
{\bf {Coherent states in $H_w$}}:
\begin{equation}\label{per-CS-w}
<\overline{w} \mid \zeta>=
 \sqrt{ \frac{(2k-1) }{\pi}}2^{2k-1}\frac{(1-\mid \zeta\mid^2)^k}
 {(w+1- \zeta
 (1-w))^{2k}}.
\end{equation}
{\bf {Coherent states in $L^2_k(\Rbig^+)$}}: The computation of
the coherent states in the space $L^2_k(\Rbig^+)$ uses the
expression (\ref{2kmy}) of the canonical basis in that space so
that (\ref{2coh-per1}) leads to:
\begin{equation}\label{1per-CS-y}
<y \mid \zeta>= \sqrt{\frac{(4\pi)^{2k}}{(2k-1)!}} (1-\mid
\zeta\mid^2)^k y^{2k-1}
 \exp(-2\pi y)
\sum_{m\geq 0} \frac{\zeta^m}{m!}P_m(y)
\end{equation}
This expression involves the generating function of the
polynomials $P_m$ which is computed from that of the Laguerre
polynomials:
\begin{equation}\label{fonc-gen}
{\cal P}(\zeta, y) \equiv \sum_{m\geq 0} \frac{\zeta^m}{m!}P_m(y)
= (1+\zeta)^{-2k} \exp\left(4\pi y\frac{\zeta}{\zeta+1}\right)
\end{equation}
The Perelomov coherent states expressed in the space
$L^2_k(\Rbig^+)$ are thus equal to :
\begin{equation}\label{Per-CS-yy}
<y\mid \zeta> = \sqrt{\frac{(4\pi)^{2k}}{(2k-1)!}}\frac{(1-\mid
\zeta\mid^2)^k }{ (1+ \zeta)^{2k}}\; y^{2k-1} \exp\left(2\pi
y\frac{\zeta-1}{\zeta+1}\right) , \; |\zeta|<1
\end{equation}

When $k=1$, these functions coincide (up to normalization) with
coherent states introduced in \cite{benedict} for the Morse
problem.

\section{Parameterization of the $SU(1,1)$ coherent states in terms of the
affine group}\label{para}

The quotient space of $SU(1,1)$ by the rotation group is a group
isomorphic to the affine group $\mathcal{A}$. This is most easily
seen when working with $SL(2,\Rbig)$ since any matrix $g$ defined
in (\ref{sl2r}) can be uniquely decomposed into the product of a
matrix of the affine subgroup by a rotation matrix as:
\begin{equation}\label{dec}
  g=\left( \begin{array}{cc}
    h_{11} & h_{12} \\
   0 & h_{11}^{-1} \
  \end{array}\right)
  \left( \begin{array}{cc}
   \cos \theta & \sin \theta \\
  -\sin \theta & \cos \theta \
  \end{array}\right)
\end{equation}
This property will now be exploited systematically.

\subsection{Affine group interpretation of Perelomov   states}
At the algebra level, the affine group generators are given in
terms of those of $SU(1,1)$ by the relations:
\begin{equation}\label{afgen}
  A \equiv J_0+J_1, \quad B \equiv J_2
\end{equation}
leading to the commutation relation:
\begin{equation}\label{comaf}
[B,A]=iA
\end{equation}
The action on the space $L^2_k(\Rbig^+ )$ is:
\begin{eqnarray}\label{aff-sur-f(y)}
A&=& 2\pi y\\ B&=& i\left( y\partial_y+1-k\right)
\end{eqnarray}
The construction of the Perelomov coherent states will now be
performed in terms of generators $J_0$ and $A,B$.\par

\noindent The use of relations (\ref{afgen}) allows to replace the
equations (\ref{1k0}) defining the fundamental state $<y|k0>$ by
the equivalent set:
\begin{equation}\label{2psi0}
  J_0<y |k0> = k <y|k0>, \quad (A-iB) <y|k 0> = k <y|k 0>
\end{equation}
Here the problem is set up in the Hilbert space $L^2_k( \Rbig^+)$
with a specific value of $k$ and the second equation, involving
the affine group generators, is sufficient to determine the
function $<y|k0>$.
 Next, we introduce the matrix   $ D(\tau,\varphi) $ of $SU(1,1)$ corresponding to the
 element  $e^{\xi J_+- \overline{\xi}J_-}$ defined in
 (\ref{exiji}):
\begin{equation}\label{dtv2}
  D(\tau,\varphi) = \left(
  \begin{array}{cc}
    \cosh (\tau/2) & -ie^{-i\varphi} \sinh (\tau/2) \\
    ie^{i\varphi} \sinh (\tau/2) & \cosh (\tau/2) \
  \end{array}
  \right)
\end{equation}
The Perelomov  coherent states are defined as displaced from
$<y|k0>$ by operator $U^k(D(\tau , \varphi))$. But since $<y|k0>$
is an eigenstate of the rotation operator, it is possible to
perform a rotation on $<y|k0>$ before applying
$U^k(D(\tau,\varphi))$ and still obtain a state belonging to the
same ray. We will take advantage of this fact to define the
coherent states by an affine transformation.
 \par
  Multiplying
$D(\tau,\varphi)$ on the right by the rotation matrix
$\Gamma_{\theta}$
  defined by the operator $e^{-i \theta J_0}$:
  %in (\ref{groupe-CS}):
\begin{equation}\label{gammarot}
  \Gamma_{\theta} = \left(
  \begin{array}{cc}
    e^{-i \theta /2} &  0 \\
    0 & e^{i\theta /2}
  \end{array}
  \right)
\end{equation}
we can determine $\theta$ (as a function of $\tau$ and $\varphi$)
so as to obtain an element of the affine group ${\cal{A}}$:
\begin{equation}\label{dmab}
D(\tau, \varphi) \Gamma_{\theta} \equiv M(a,b)
\end{equation}
where $M(a,b)$ is defined in (\ref{mab}).  This relation gives a
one-to-one correspondence between parameters $(\tau, \varphi )$
and  $(a,b)$. Using definition (\ref{zeta}) of $\zeta$ in terms of
$(\tau , \varphi )$ leads to the expressions:
\begin{equation} \label{zetaab}
\zeta = \frac{1-a +ib}{1+a -ib}\quad , \quad e^{-i\theta} =
  \frac{1+a +ib}{1+a -ib} \label{2ident-2}
\end{equation}
Coherent states $<y|ab>$ are now defined as transforms of $<y|k0>$
by the operator $U_k(M(a,b))$. Their explicit form   is obtained
using (\ref{repu}) and (\ref{k0-y}):
\begin{equation}\label{afcs}
<y\mid a b> \equiv U^k(M(a,b)) <y|k 0> =
\frac{(4\pi)^k}{\sqrt{(2k-1)!}} a^{k} \exp(2\pi
 (-a+i b)y)\, y^{2k-1}
\end{equation}
 They are related  to Perelomov states by:
\begin{equation}\label{ident-3}
\mid a b > = \left( \frac{1+\zeta }{1+ \overline{\zeta}} \right)^k
\mid \zeta >
\end{equation}
where the expressions of the parameters  $(a,b)$ in terms of
$\zeta$ are deduced from (\ref{zetaab}).\par

  The
completeness
 relation in variables $(a,b)$ is obtained from
(\ref{rel-fer-zeta}) and reads:
\begin{equation}\label{compa}
\frac{2k-1}{4\pi} \int _0^\infty \frac{da}{ a^{2}}\int
_{-\infty}^\infty db \mid a b
>< a b \mid =1
\end{equation}

 As a result, the Perelomov coherent states can be characterized either
 by $(a,b)$ or $ \zeta$. They form an overcomplete basis of the
space $L^2_k(\Rbig^+)$ that is constructed from the fundamental
state $<y|k0>$ by applying an affine group transformation.
\subsection{Conditions for the  affine group wavelets to be $SU(1,1)$ coherent states}

General coherent states associated with the affine group, also
known as wavelets, may be constructed using fundamental states
different from $|k0>$ \cite{klauder}, \cite{kaiser}. In fact,
choosing an element $\sigma_0(y) \in L^2_k(\Rbig^+)$ ("mother
wavelet"), we can construct the family $\sigma_{a,b}(y)$ by
application of the representation (\ref{repu}) of group
${\cal{A}}$ as:
\begin{equation}\label{sigmab}
  \sigma_{ab} (y) \equiv a^{1-k} e^{2i\pi by} \sigma_0(ay)
\end{equation}
Any state $\psi (y)$ belonging to $L^2_k(\Rbig^+)$ can be
developed on the family $\{ \sigma_{ab}(y) \}$ with coefficients
("wavelet coefficients") given by:
\begin{equation}\label{wavelet}
  C(a,b) = (\psi (y) , \sigma _{ab} (y) )
\end{equation}
\noindent where $( \; , \; )$ denotes the scalar product
(\ref{scay}). A direct computation shows that the state $\psi (y)$
can be reconstructed from its coefficients $C(a,b)$ provided the
mother wavelet $\sigma_0(y)$ satisfies the condition:
\begin{equation}\label{admis}
\int_0^{\infty} |\sigma_0 (y) |^2 y^{-2k} \; dy < \infty
\end{equation}
This so-called admissibility condition is usually written with
$k=1/2$. Recall that it is possible to choose a particular value
of $k$ when considering the affine group alone because of the
equivalence of its representations for different values of
$k$.\par

\noindent Thus there exists an infinite family of overcomplete
bases constructed with the affine group for space
$L^2_k(\Rbig^+)$. However, if the invariance (up to a phase) by
$SU(1,1)$ is required, the basic state $\sigma_0 (y)$ must be an
eigenstate of the rotation operator $J_0$. This restricts the
choice to $\sigma_0 (y) = <y|km>$.

\subsection{Morse coherent states}

Group theoretical arguments have led to use the $SU(1,1)$ coherent
states in the problem of the Morse oscillator \cite{gerry}. But
 several different representations are then required for a
complete description. A more satisfactory family of coherent
states has been introduced as eigenstates of an annihilation
operator by Benedict and Molnar in \cite{benedict} and shown to be
related to affine coherent states \cite{benedict2}. The present
study allows us to find the exact relation between the two sets.
 \par

  The Morse potential considered in  \cite {benedict}  has the form:
\begin{equation}\label{morse}
  V(x) = (s+\frac{1}{2} - e^{-x})^2
\end{equation}
where $s$ is a real parameter such that $s>1/2$. The relevant
Hilbert space of the problem is $L^2_k(\Rbig^+)$ for $k=1$ and the
system of coherent states can be constructed by displacement of
the fundamental state:
\begin{equation}\label{csb}
  \phi_0 (y) = {\displaystyle \frac{(4\pi )^s}{\sqrt{\Gamma
  (2s)}}\,
  y^{s} \, e^{- 2\pi y}}
\end{equation}

When $s=1$,
 this state    coincides with the state $<y|k0>$ defined in
  (\ref{k0-y}) and the corresponding coherent states are identical with
  Perelomov states $|\zeta>$.
  %\begin{equation}
%|\beta > = | -\zeta >
%\end{equation}
When $s\neq 1$, the state (\ref{csb}) is no longer invariant by
the subgroup of rotations but, as recalled in section 5.2, it can
still be used to construct affine coherent states. The latter
  are, up to a phase, equal to the states considered in
\cite{benedict}.\par

In conclusion, Perelomov coherent states for the discrete series
representation of $SU(1,1)$ are a subset of the coherent states
considered in \cite{benedict} for a Morse potential problem.\par

\section{Consequences of the new characterization of the $SU(1,1)$ coherent states}\label{prop}
   The properties of coherent states obtained by applying
 the displacement operator on a fiducial state  come directly from
 those of the latter. In particular, when
 the fiducial state is $|k0>$, the corresponding coherent states
 have minimal properties and satisfy   equations derived from (\ref{1k0}). The explicit
  results  are
 most easily derived using the parameterization in terms of the
 affine group, as shown below.
\par

Let  $O_1$ and $O_2$ be two self-adjoint operators and let
$<O_i>$, $i=1,2$
  denote their  mean values
 in an arbitrary state   $\mid \psi >$. Introduce the centered operator $\ov{O}_i$
 as:
\begin{equation}\label{oi}
  \ov{O}_i = O_i - <O_i>
\end{equation}
and define the mean square deviations:
\begin{equation}\label{msd}
  \Delta_i =<\ov{O}_i^2>
\end{equation}
and the correlation:
\begin{equation}\label{cor}
  \Delta_{12} = <\ov{O}_1 \ov{O}_2 + \ov{O}_2 \ov{O}_1 >
\end{equation}
Writing that the norm of the state $(\ov{O}_1+ i\lambda
\ov{O}_2)\mid \psi>$
 is positive for every complex value of $\lambda$ leads to the generalized uncertainty relations:
 \begin{equation}\label{rel-inc1} 4 \Delta_1 \Delta_2 -
 \Delta_{12}^2
\geq (<i[O_1,O_2]>)^2
\end{equation}

Starting from a real $\lambda$, one obtains the more usual
relation:
\begin{equation}\label{rel-inc2}
4 \Delta_1 \Delta_2 \geq (<i[O_1,O_2]>)^2
\end{equation}

The equality in (\ref{rel-inc1}) is obtained for states $|\psi>$
verifying:
\begin{equation}\label{minlam}
 (\ov{O}_1+ i\lambda \ov{O}_2)\mid \psi> = 0
\end{equation}
for complex values of $\lambda$. When $\lambda$ is real, the
correlation $\Delta_{12}$ vanishes and the corresponding state
minimizes the stricter relation (\ref{rel-inc2}).
\par

This general scheme is now applied to the generators $A$, $B$ of
the affine group and to the coherent states $|ab>$.

Property (\ref{2psi0}) implies that the affine coherent states
verify the following relation:

\begin{equation}\label{pro-aff-CS}
M(a, b)(A-iB) M(a, b)^{-1}\mid a,b> \equiv ((a-ib)A-iB)\mid a,b> =
k \mid a,b>
\end{equation}
which allows to compute the mean values of the affine generators:

\begin{equation}\label{a}
<A> = k a^{-1}, \quad <B> = -k b a^{-1}
\end{equation}

\noindent Relations (\ref{pro-aff-CS}) and (\ref{a}) lead to the
equations characterizing the coherent states $\mid a b >$:
\begin{equation}\label{annihab}
\left(\overline{B}+i(a-ib)\overline{A} \right)\mid a b > = 0
\end{equation}
This equation is of the form (\ref{minlam}). In the present case,
  the parameter $\lambda$ has a definite value $\lambda = a-ib$
depending on the state $|ab>$. The correlation $\Delta_{12}$
between $A$ and $B$ vanishes only for $b=0$. However, for each
state, it is possible to introduce uncorrelated operators $A$ and
$a^{-1}(B+bA)$.
\par
The exploitation of these results will be performed in terms of
$\zeta$. The equation (\ref{pro-aff-CS}) becomes~:
\begin{equation}\label{pro3-PCS}
\left((1 -\zeta) J_0 - \zeta J_+ + J_- - k(\zeta+1)\right)\mid
\zeta>= 0
\end{equation}

 Let us denote $\tilde{J_i} \equiv \exp(\xi
J_+-\overline{\xi}J_-)J_i \exp(-\xi J_++\overline{\xi}J_-)$. The
coherent states verify two relations resulting from the properties
of the fundamental state (\ref{1k0}):
\begin{equation}\label{pro1-PCS}
\tilde{J_0}\mid \zeta> \equiv \vec{n}.\vec{J} \mid \zeta>= k \mid
\zeta>
\end{equation}
and
\begin{equation}\label{pro2-PCS}
\tilde{J_-}\mid \zeta> \equiv \left(J_--2 \zeta J_0+ \zeta^2J_+
\right)\mid \zeta>= 0.
\end{equation}
The combination of (\ref{pro2-PCS}) and (\ref{pro3-PCS}) gives two
simpler equations~:
\begin{equation}\label{pro4-PCS}
\left(J_0 - \zeta J_+ -k\right)\mid \zeta>= 0
\end{equation}
and
\begin{equation}\label{pro5-PCS}
\left(\zeta J_0 - J_- +k \zeta\right)\mid \zeta>= 0
\end{equation}

To compute the mean values of $<\vec{J}>$, we
 multiply these equations on the left by $< \zeta \mid $ and find:

\begin{equation}
\label{mean-J}
\begin{array} {cll}
<J_0>& = \; k {\displaystyle
\frac{1+\mid\zeta\mid^2}{1-\mid\zeta\mid^2}} & = \;
k\cosh\tau\\[3mm] <J_1>& = \; k{\displaystyle
\frac{\zeta+\overline{\zeta}}{1-\mid\zeta\mid^2} }&= \;
k\sinh\tau\cos\varphi\\[3mm] <J_2>& = \; i k{\displaystyle
\frac{\zeta-\overline{\zeta}}{1-\mid\zeta\mid^2}}  & = \;
k\sinh\tau\sin\varphi
\end{array}
  \end{equation}
Thus the two vectors $<\vec{J}>$ and $\vec{n}$, defined in
(\ref{hyp}), have the same direction~:
\begin{equation}\label{pro44-PCS}
<\vec{J}> = k \vec{n}
\end{equation}

\noindent Due to equations (\ref{pro4-PCS}), (\ref{pro5-PCS}) and
to results (\ref{mean-J}), the coherent states $\mid \zeta>$
verify the following equations
\begin{eqnarray}\label{annihzeta}
\left(\overline{J_1}+i\frac{\zeta^2+1}{\zeta^2-1}\overline{J_2}\right)\mid
\zeta> &=&0\\
\left(\overline{J_0}-\frac{2\zeta}{\zeta^2+1}\overline{J_1}\right)\mid
\zeta> &=&0\\
\left(\overline{J_0}+i\frac{2\zeta}{\zeta^2-1}\overline{J_2}\right)\mid
\zeta> &=&0\\
\end{eqnarray}

\noindent The interpretation of these equations for $\zeta$ real
shows that the coherent states $|\zeta>$ minimize the usual
 uncertainty relation (\ref{rel-inc2}) for
the pairs $(J_1, J_2)$ and $(J_0, J_2)$. These states are
associated with the section of the upper sheet of the hyperboloid
by the plane $(n_0,n_1)$. Similarly, the section by the plane
$(n_0,n_2)$ corresponds to coherent states $|\zeta>$ with $ \zeta$
purely imaginary that minimize the relation (\ref{rel-inc2}) for
the pairs $(J_1,J_2)$ and $(J_0,J_1)$.\par

For other values of $\zeta$, the operators $(J_i,J_j)$ are
correlated. However, it is always possible to construct
uncorrelated operators in the form $J_i+\lambda J_j$ and $J_i +
\mu J_j$, where $\lambda$ and $\mu$ depend on the parameter
$\zeta$.

\section{Conclusion}

 Realizing the discrete series representation  of $SU(1,1)$ labeled by $k$    and the
corresponding representation  of the affine group  in the same
Hilbert space,
   we have been able to make a precise comparison of the coherent states attached
   to the two groups. The Perelomov coherent states constructed either on $|k0>$
   or $|km>$ have been found identical,  up  to a phase, to special families of
   affine coherent states or wavelets. Conversely,
affine group coherent states obtained from a basic state that is
invariant under rotations form an overcomplete basis that is
invariant as a whole under the $SU(1,1)$ representation.\par

The characterization of the rays in terms of affine wavelets has
several advantages: Minimal properties of the states and
characteristic equations are easily obtained. More fundamental is
the result that the set of $SU(1,1)$ coherent states is strictly
invariant by action of the affine group representation, while it
is invariant only up to a phase by action of $SU(1,1)$. These
results stress the importance of determining the invariance group
of a problem to be able to take full advantage of the properties
of the system of coherent states.

%%%%%%%%%%%

\end{document}